**COLLABORATORY AGAINST HATE** | **RESEARCH AND ACTION CENTER**

# "I don't trust them": Exploring Perceptions of Fact-checking Entities for Flagging Online Misinformation

The spread of misinformation through online social media platforms has had substantial societal consequences. As a result, platforms have introduced measures to alert users of news content that may be misleading or contain inaccuracies as a means to discourage them from sharing it. These interventions sometimes cite external sources, such as fact-checking organizations and news outlets, for providing assessments related to the accuracy of the content. However, it is unclear whether users trust the assessments provided by these entities and whether perceptions vary across different topics of news. We conducted an online study with 655 US participants to explore user perceptions of eight categories of fact-checking entities across two misinformation topics, as well as factors that may impact users' perceptions. We found that participants' opinions regarding the trustworthiness and bias of the entities varied greatly, aligning largely with their political preference. However, just the presence of a fact-checking label appeared to discourage participants from sharing the headlines studied. Our results hint at the need for further exploring fact-checking entities that may be perceived as neutral, as well as the potential for incorporating multiple assessments in such labels.

**Hana Habib**
Carnegie Mellon University

**Sara Elsharawy**
Independent Researcher

**Rifat Rahman**
Bangladesh University of Engineering and Technology

The Collaboratory Against Hate: Research and Action Center at Carnegie Mellon University and the University of Pittsburgh aspires to develop and support innovative multidisciplinary, interdisciplinary, and cross-university research aimed at understanding how extremist hate is generated, how it circulates in online and real-life spaces, and how it polarizes society and provokes harmful and illegal acts, especially toward communities of color and other minoritized groups. We seek to develop effective interventions to inhibit every stage in the creation and growth of extremist hate groups and to minimize their destructive consequences.

www.collabagainsthate.org

## 1. Introduction

With the rise of social media, misinformation — misleading or inaccurate news — has surged. Social media algorithms often amplify this over accurate news, and platform designs enable rapid, unreflective sharing [25]. Misinformation has influenced events like Brexit, the US 2016 election, and the January 6, 2021 US Capitol attack [31, 21]. During the COVID-19 pandemic, it caused confusion about health guidelines [9] and fueled prejudice, such as false claims on WhatsApp inciting mob violence and scapegoating of Muslims [29].

Though still incomplete [31], major platforms like Facebook and Twitter[1] have taken steps to combat misinformation. They employ prebunking, offering informative messages against false narratives, and direct users to reliable sources [11, 17]. Content labels also warn users about potential inaccuracies, often leveraging external fact-checkers and news sources, e.g., [23]. Past work has primarily focused on the media source of the news when assessing the effectiveness of content labels, e.g., [8, 13, 18], finding mixed results related to perceptions of the publisher of the content. Questions remain about users' trust in different entities when referenced as fact-checking sources, rather than publishers of the news content, and if certain fact-checking entities[2] are more effective than others in dissuading users from sharing misleading content. The consistency in perceived trustworthiness across different news topics is also unclear; for example users may have greater

---

[1] Twitter, now X, has changed its approach to addressing misinformation, now largely relying on user-contributed Community Notes.

[2] In the context of our study, "fact-checking entities" refer to organizations that may or may not currently assess the accuracy of news content but could hypothetically perform such assessments.

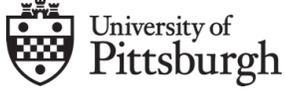

trust in government agencies for health-related information than other types of news.

To address this gap, we examined perceptions of eight categories of fact-checking entities through an online study with 655 US participants. Participants were presented with screenshots of news headlines stylized as posts on a social media platform that were flagged for inaccuracies with a content label that referenced a fact-checking entity. We asked participants to rate their likelihood of sharing the headline, the helpfulness of the accuracy assessment, and the trustworthiness of the particular fact-checking entity referenced in the content label. To evaluate whether perceptions of different categories of fact-checkers were consistent across different topics of misinformation, each participant evaluated one COVID-19 vaccine-related headline and another with anti-Muslim sentiment. In addition to their perceptions about the headline, participants provided feedback on trusted news sources, distrustful ones, and their personal strategies for verifying potentially inaccurate news.

Our study showed that participants' views on fact-checking entities varied widely, particularly along political lines, with conservatives expressing more negative perceptions than liberals. Major news outlets were often seen as biased. However, perceptions of fact-checking organizations like Poynter.org, were not found to vary with political preference or news topics. Participants reported to cross-check suspicious news across multiple sources before deciding on its validity. Our findings suggest misinformation interventions should reference multiple fact-checking sources rather than just one to avoid perceptions of bias in the fact-checking assessment.

## 2. Background & Related Work

The emergence of social media has led to a significant reduction in the cost of sharing information, making news access and dissemination more accessible. This digital shift has allowed the proliferation and polarization of news content and information sources, representing a wide spectrum of ideological and cultural perspectives. Such views expose the diversity of online audiences, in addition to the various factors that shape their online interactions with news content, especially with respect to sharing behavior.

### 2.1 Perceptions of Information Sources

Previous research explored users' perception of news sources online and how they can impact their sharing behavior. It has been shown that users can give precedence to trusted sources over ideological-driven ones [5]. While some studies have previously shown that people viewed the news as more credible when shared by a media source that they trust [13, 18], others have found that the news publisher had no significant impact on whether participants perceived a headline as accurate or expressed an intention to share it [8]. Further, previous work has found that even when presented with correct information, false beliefs continue to persist [7, 10, 32] highlighting how users' perceptions can impact the efficacy of these labels. While information sources remain a significant variable in online news consumption, there still remains a literature gap that explores the perception of entities as fact-checkers rather than sources of news.

### 2.2 Social Media Design & Interventions

Design of social media platforms also play a role in the spread of misinformation. The common "share" feature allows users to re-share news posts with a single click, potentially leading to a viral spread [33]. Prior research argued that misinformation interventions should prioritize the reduction of sharing prompts as sharing behaviors were observed to continue despite corrective measures [28].

Online platforms have used various methods to alert users to potential misinformation. Among the most common interventions have been social media content warning labels [12]. For example, Facebook's warning labels highlight the fact-checking audit of the news article, name of the fact-checking entity, and provide a link for additional context [22]. Twitter also uses prebunking to preemptively provide users with accurate information to counter false narratives [36]. YouTube's labels of state-media funded content appear below streamed videos to inform the user of a media channel's government connection [37]. Other interface displays include social media metrics which indicate content popularity or appeal by the user's social network [4].

### 2.3 Evaluations of Misinformation Interventions

There has been some prior evaluation of different misinformation interventions. Twitter's internal research found that their label design [35] led to an increase of users clicking on prebunking information [2]. A study by Bradshaw et al. [3] explored the effectiveness of YouTube's labels, finding that users' existing beliefs and opinions should be considered when presenting interventions to mitigate misinformation. Researchers also discovered that the efficacy of the source notice label is dependent on the interface factors that help users perceive it as intended [26].

To complement this prior work, this study further investigates the effect of different types of entities when presented as fact-checkers, rather than as a source, on users' perception and sharing behavior of online news content.

## 3. Methods

We conducted an exploratory online study to investigate perceptions of fact-checking entities and the consistency of these perceptions across two topics of misinformation.

### 3.1 Study Design

Participants were asked to evaluate two news headlines, building on methods used in prior misinformation work (e.g., [6, 17]). Headlines were styled as social media posts with a content label warning, inspired by a design used by Facebook, stating the "independent fact-checkers say this information has some factual inaccuracies," followed by the name of the fact-checking entity. We evaluated eleven entities corresponding to eight different categories of organizations that could potentially serve in a fact-checking role. To facilitate comparison of perceptions across topics, each participant was assigned two different headlines with two different entities at random, with one headline portraying anti-Muslim sentiment and one skepticism of COVID-19 vaccination. Figure 1 contains examples of screenshots shown to participants.

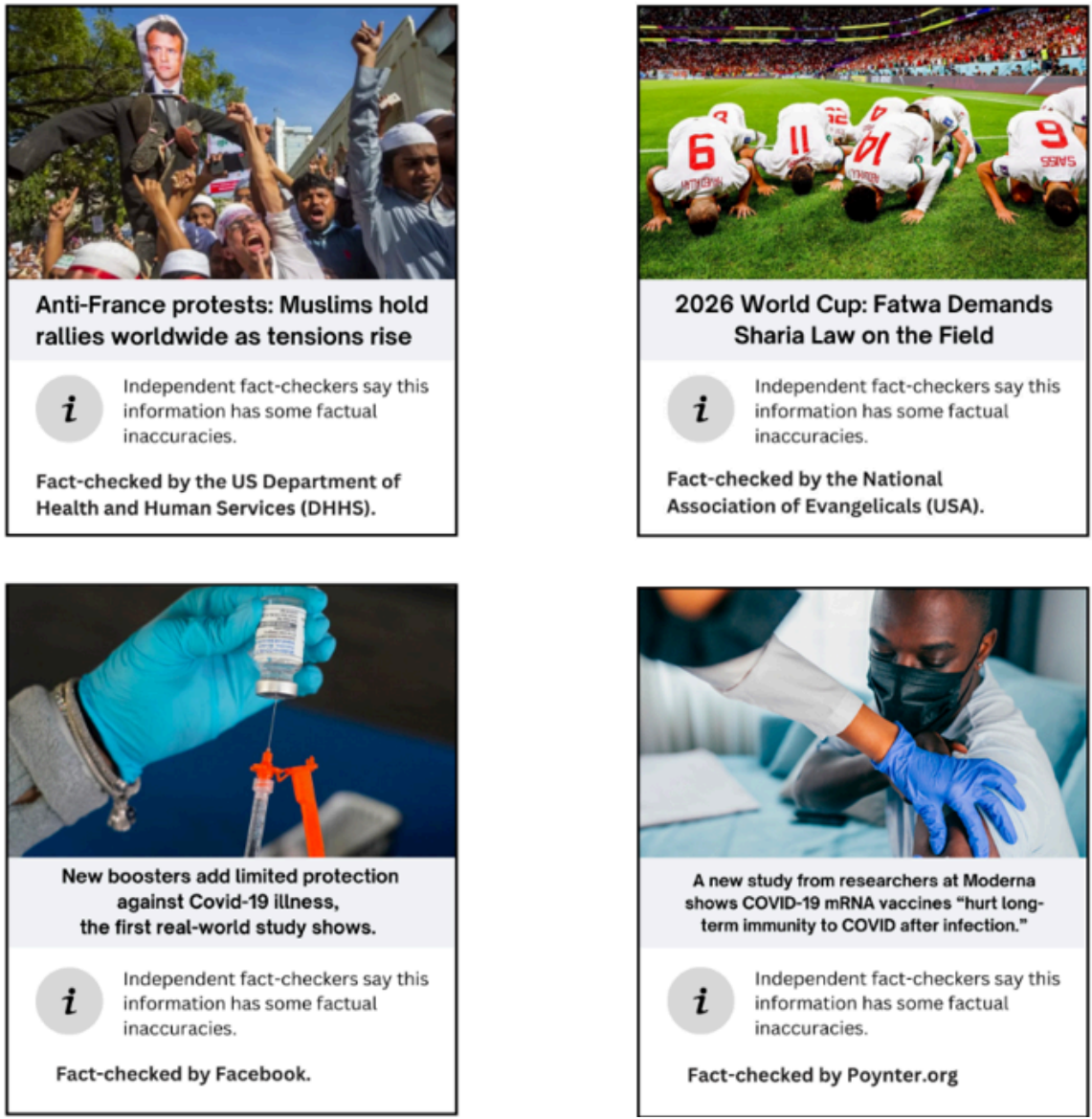

Figure 1: Examples screenshots of content labels shown to participants of the four news headlines with a corresponding image

### 3.1.1 Selection of News Headlines

We included four headlines in our study, shown in the examples provided in Figure 1. We chose headlines that reflected major recent global events with which participants would have some degree of familiarity. We wanted to isolate participant perceptions to the specific entity providing the fact-checking assessment, thus we chose not to include the publisher of the article content as part of the content label shown to participants. Further, following

**COLLABORATORY AGAINST HATE** | **RESEARCH AND ACTION CENTER**

| Categories | Fact-checking Entities | |
|---|---|---|
| | *Topic: COVID-19 Vaccination* | *Topic: Anti-Muslim sentiment* |
| Government | Center for Disease Control (CDC) | Department of Health and Human Services |
| Religious Orgs | National Association of Evangelicals (NAE) | |
| Universities | The Collaboratory Against Hate - Carnegie Mellon University & The University of Pittsburgh | |
| NGOs | Doctors without Borders | Amnesty International |
| News Entities | CNN | |
| Fact-checking Orgs | Poynter.org | |
| International Orgs | World Health Organization (WHO) | UN Human Rights Council (HRC) |
| Companies | Facebook | |

Table 1: Overview of the eight categories and 13 entities selected for the study.

recommendations by Pennycook et al. [27], we came up with two headlines that were factually true and two that contained inaccuracies (one for each topic) to avoid the veracity of the headline from playing a confounding factor. We also tried to make the headlines ambiguous in their veracity so that participants relied on the fact-checking entities for their assessment, rather than their previous knowledge.

The factually accurate headlines we generated were inspired by previously published headlines from reputable mainstream news outlets, such as the BBC and CNN, which portrayed Muslims or COVID-19 vaccination in a negative light. We sourced ideas for headlines containing inaccuracies by referring to fact-checking organizations, such as Poynter.org, and exploring the dialogue around global events, such as the World Cup, on social media platforms.

### 3.1.2 Selection of Fact-checking Entities

Our study included eight categories of entities (Table 1) that we thought would be interesting to explore as the role of fact-checkers. We did not want to limit our selection of entities to only those that currently have fact-checking capabilities, but rather include those that may engender some level of trust in social media users and could hypothetically serve in a fact-checking role. We drew inspiration for categories and entities to include from the 2017 US results of the World Values Survey (WVS) [34] which reported on participants' confidence level in certain categories of organizations (reported in Table 4 in Appendix B). The majority of WVS US participants were found to have "a great deal" or "quite a lot" of confidence in religious organizations, universities, NGOs, and international organizations (such as World Health Organization), thus we included these categories/entities in the study.[3] Considering the prominence of Christian Evangelicals in American political and cultural discourse, we chose the National Association of Evangelicals (NAE)[4] as the religious organization in our evaluation, hypothesizing that at least a subset of participants — particularly those that were conservative-leaning — would have favorable opinions of the organization. Though WVS participants were found to lack confidence in government and news entities, we thought it made sense to include entities corresponding to those categories given the context of the study and their potential capabilities to serve as fact-checkers. Similarly, we included a non-profit that has a dedicated fact-checking mission (Poynter.org) as such fact-checking organizations are commonly referenced in misinformation interventions. When choosing an entity for each category, we aimed to identify those that would have familiarity to our survey population of US participants. We also selected topic-specific entities for the government, NGO, and international organization categories to better align the main work of the entity with the news topics being studied. The choice of the news entity and fact-checking organization (i.e., CNN and Poynter, respectively) was made by referring to the Tranco rankings[5] of possible options within that category and selecting the entity that had the highest web traffic. We also included Facebook in our list of entities, given its prominence as a social media platform and Meta's role in facilitating fact-checking.

---

[3] For the university entity we chose our affiliated research center: "The Collaboratory Against Hate - Carnegie Mellon University & The University of Pittsburgh"

[4] NAE: https://www.nae.org

### 3.1.3 Survey Questions

Participants first answered questions specific to the headline and entity displayed. We asked them to rate the helpfulness of the assessment provided, their likelihood of sharing the content as measured in prior work exploring misinformation, e.g., [8], and trustworthiness of the entity in assessing or presenting news on a 5-point Likert scale with the options "not at all", "slightly", "somewhat", "moderately", and "extremely". Participants were also asked to provide supporting details behind their helpfulness and sharing intentions ratings. We also asked participants whether they felt that the entity deals fairly with all sides, or if it tends to favor one side. Then, we measured whether the participants were familiar with the entity and headline.

After entity-specific questions, participants were requested to elaborate on how they judge news accuracy when they are not sure about the veracity of a claim. Participants were then asked the WVS confidence question for the categories included in the study. Finally, we asked participants to describe what made them trust an entity (e.g., specific person, news outlet) to provide accurate information, as well as about entities they do not trust.

We also gathered participants' demographic information including age, gender, minoritized status, education & income levels. Given the partisan nature of news dissemination, we also collected information reflecting participants' political preferences and engagement with politics and public affairs. To better control for participants' pre-existing beliefs regarding COVID-19 and Muslims, we asked participants their level of belief in five statements related to these topics that have been used in a similar manner in prior work [19, 16]. The text of our survey questions is included in Appendix A.

### 3.2 Data Collection

We recruited 655 US-based participants from Prolific to take part in our study. Participants were required to be older than 18 years old, fluent in English, and completed an online consent form prior to answering survey questions. Participants were not required to be social media users to complete the study. The median completion time was 10 minutes and participants received $2.50 in compensation.

### 3.2.1 Ethical Considerations

Participants were informed in the consent form that they might be exposed to content that might seem offensive or misleading, although any misinformation would be flagged as such as part of the study design. Further, we debriefed participants at the end of the survey to make them aware of the headlines that were actually factually true to prevent misconceptions about factual news. We also disclosed that the images they viewed were manipulated for the study and not all the entities included actually perform fact-checking.

No personally identifiable or confidential information (that could track the activity of the participants) was collected. The study data was accessible to only members of the study team through Google Drive and Qualtrics.The study data was accessible to only members of the study team through Google Drive and Qualtrics. The IRB at our respective institutions reviewed our protocol and determined it exempt from full review.

### 3.2.2 Experimental Procedure

Our survey was implemented in Qualtrics where participants first completed the consent form and indicated their eligibility. Participants were instructed: "Imagine that you saw the following posts on a social media platform. Please answer the following questions with regards to the displayed posts. Note that you should not reveal any private or personally-identifiable information about yourself OR others in answers to your open-ended questions." They were then presented with an image of a social media post that

---

**COLLABORATORY AGAINST HATE** | **RESEARCH AND ACTION CENTER**

contains a news headline, corresponding photo, an alert that expresses some factual inaccuracies in the given headline, and the fact-checking entity that checked the news. Figure 1 shows the image format for four different headlines with randomly chosen four fact-checking entities based on the topic. We did not show the whole article corresponding to the headline as we thought it would divert the focus of the participants to evaluate news content rather than fact-checking entities [17]. After observing the image format of the headline, participants answered the entity-specific questions.

To avoid fatigue, participants were asked to evaluate only two of the four headlines included in the study. Participants were assigned one headline per topic so that we could evaluate the consistency of participants' perceptions of entities across topics. To mitigate order effects, the order of the topics presented was randomized. Participants were also assigned two different entities across topics. Assignment of factually true and false headlines were done at random, with it possible for a participant to see both factually true or factually false headlines. Figure 2 provides an overview of the experimental procedure.

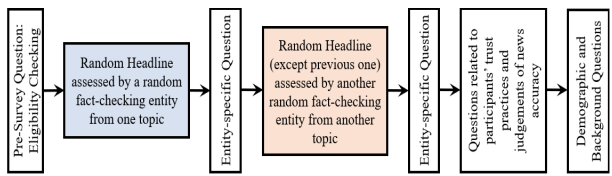

Figure 2: Overview of the experimental steps of the online survey.

Before distributing our survey study, we conducted two rounds of pilots with 40 participants to test our study implementation. We also used the pilot to better estimate the duration and adjusted the compensation rate accordingly. We launched our pilots at two different times to capture a diverse set of participants. After our pilots, we identified a minor bug in the survey's Javascript and

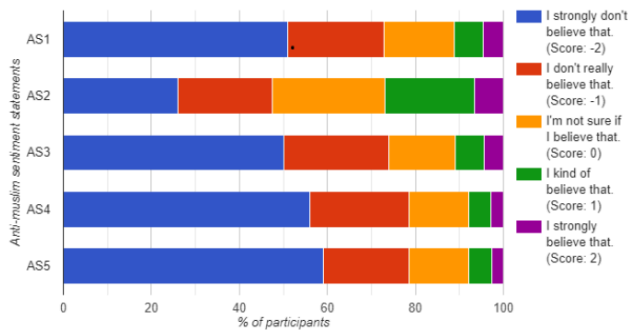

(a) Participants' level of belief in the statements (AS1-AS5) expressing anti-Muslim sentiment.

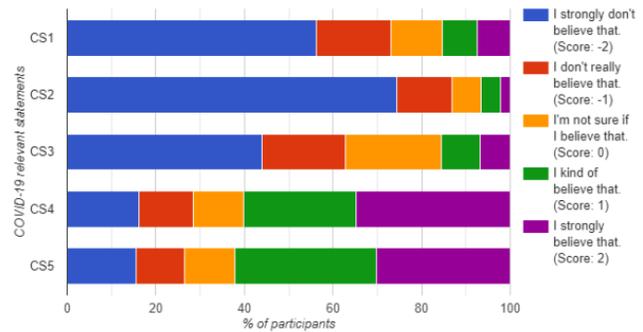

(b) Participants' level of belief in the statements (CS1-CS5) relevant to COVID-19.

Figure 3: Participants' sentiment and pre-existing belief towards anti-Muslim sentiment and COVID-19 related statements in a 5-point likert scale ranging from -2 to 2

addressed it. Our final data collection was done over six rounds in August 2023. We used Prolific's gender-balanced option to avoid a gender skew in our sample. Further, considering the relationship between political affiliation and beliefs about fake news observed in prior work [24], we utilized available screening filters related to political preference to balance our sample between conservatives, moderates, and liberals.

## 3.3 Data Analysis

During data cleaning, we removed a duplicate response from one participant and kept their first response. We suspected use of an AI tool to answer free response questions by another participant, as their responses were much longer than the average but their duration on the survey was about average, and removed their response from analysis. Therefore we analyzed data from 654 participants.

Both qualitative and quantitative approaches were performed for analyzing data. We conducted a thematic analysis of the qualitative responses from the survey. One author reviewed a subset of the responses and identified a preliminary codebook. The initial codebook was discussed and collaboratively refined with another author. The data was then coded by one of these two authors. The final codebooks used in the analysis are provided in Appendix C, D, and E.

We applied statistical tests to compare the participants' ratings for helpfulness, trustworthiness, sharing intentions, and fairness. For such analyses, participants' five-scale Likert responses were binned into three categories ("agree", "neither agree or disagree", and "disagree") to account for zero counts. To measure the significant differences among independent groups (e.g., participants' ratings on fact-checking entities) or subgroups (e.g., age/gender/political preferences-based subgroup), we relied on Pearson's Chi-squared tests of independence [30] as the dependent variables were categorical and there was one independent variable with two or more levels in each test. We considered using a significance threshold of $\alpha$=0.05 for these tests.

## 4. Participants

Our study included data from 654 US-based participants whose demographics are reported in Table 2. Our study sample was balanced across genders, with 49% identifying as male and 49% identifying as female (the remainder identified as non-binary, self-described, or preferred not to answer). The mean age of participants was 40 years old (median: 36). Participants varied in their reported household income and educational attainment, with 49% reporting to have a Bachelor's degree or higher. Nearly a quarter (23%) of participants reported belonging to a minority ethnic or religious group in the US. Nearly all participants reported to be at least somewhat engaged with politics and public affairs, with 46% responding that they paid attention to these topics at least sometimes or about half of the time and 51% most or all of the time. Our sample was balanced in terms of political preference as 44% reported to be somewhat or strongly conservative, 15% as moderates, and 40% somewhat or strongly liberal.

| Gender | | Age (Year) | | Minoritized Status | | Education | | Income | | Political Preference | |
|---|---|---|---|---|---|---|---|---|---|---|---|
| Female | 48.9% | Mean | 36 | No | 74.0% | High School or Less | 16.8% | <$30k | 20.8% | Strongly conservative | 17.1% |
| Male | 48.9% | Med. | 40 | Yes | 22.8% | Some College | 22.2% | $30k - $59,999 | 27.1% | Somewhat conservative | 26.6% |
| Non-bin ary/ third gender | 1.4% | Std | 14.1 | NR | 3.2% | Associates/ Bachelors | 47.1% | $60k - $99,999 | 27.5% | Moderate | 14.5% |
| | | Min | 18 | | | | | ≥$100k | 21.9% | | |
| Self-des cribed | 0.15% | Max | 76 | | | Graduate/ Professional | 13.0% | NR | 2.8% | Somewhat liberal | 19.7% |
| | | NR | 0.46% | | | | | | | Strongly liberal | 20.6% |
| NR | 0.61% | | | | | NR | 0.92% | | | NR | 1.4% |

Table 2: Summary of participant demographics. Those who reported belonging to an ethnic or religious minority group in their country of residence were counted as "yes" for minoritized status. NR = no response

Figure 3 depicts participants' beliefs in statements related to Muslims and COVID-19. We observed that while the majority of the participants did not believe in the false statements related to COVID-19 (CS1, CS2, and CS3) and did believe in the factually correct statements (CS4 and CS5), approximately 15-25% seemed to hold misconceptions about the topic. However, participants' average across the five statements (AS1-AS5) to measure anti-Muslim sentiment was -1.0 ("I don't really believe that") with a median and standard deviation -1 and 1.17, respectively, indicating a general lack of preexisting bias against Muslims.

**COLLABORATORY AGAINST HATE** | **RESEARCH AND ACTION CENTER**

## 5. Results

Our results found that participants' perceptions of the entities studied varied greatly, according to age, political preference, and familiarity. Participants trusted entities that they perceived to be unbiased and distrusted those they thought were one-sided in their reporting. However, content labels do appear to have some impact on participants' likeliness to share misinformation.

### 5.1 Helpfulness of the Assessment

Participants varied in their responses as to how helpful they found the accuracy assessment by the fact-checking entity referenced to be, with 21% of responses rating it to be "not at all" helpful, 40% as "slightly" or "somewhat", and 39% as "moderately" or "extremely helpful". Helpfulness ratings did not vary significantly for the entities referenced in the headlines portraying anti-Muslim sentiment but did vary significantly for COVID-19 headlines (p< 0.001) with the WHO receiving the most "moderately" or "extremely" helpful ratings and the NAE receiving the most "not at all" helpful ratings.

While ratings did not significantly differ between males and females, participants' ratings did vary significantly for some of the entities evaluated according to age, political preference, and familiarity. Across age groups, Facebook (p= 0.05) and the UN Human Rights Council (p= 0.007) had significant differences for anti-Muslim headlines and CNN (p= 0.02) and WHO (p= 0.02) had significant differences for COVID-19 headlines. Participants older than 45 rated the assessments by Facebook and the UN Human Rights Council less favorably than younger age groups, while those 18-30 had more favorable ratings for CNN and WHO compared to older participants. All entities except Poynter.org for anti-Muslim headlines and NAE, Poynter.org, and Facebook for COVID-19 significantly varied with political preference. Except for NAE, all were rated less positively by conservative-leaning participants than liberal-leaning participants. Familiarity also impacted the helpfulness ratings of the UN Human Rights Council (anti-Muslim), Amnesty International (anti-Muslim), and WHO (COVID-19). Those unfamiliar with the UN Human Rights Council and Amnesty International significantly more likely to rate them as "not at all" helpful, while the converse was true for WHO.

Of the 223 responses that rated the assessment provided by the fact-checking entity for a headline as "not at all" helpful, 117 expressed negative sentiment toward the fact-checking entity with the National Association of Evangelicals being mentioned most frequently in these responses. In the 526 responses that rated the helpfulness as "slightly" or "somewhat," participants most commonly mentioned that there was not enough information provided concerning the inaccuracies or the fact-checking entity (168 responses). For example, one participant noted "It doesn't tell me what's inaccurate about the title, and anything about the fact-checking organization." Those who found the assessment "moderately" or "extremely" helpful (504 responses) most frequently mentioned the importance of fact-checking generally (224 responses) or expressed a positive view of the fact-checking entity presented (159 responses). The WHO, CDC, and Doctors Without Borders were most frequently mentioned in these responses. A detailed summary of participant responses is provided in Appendix G.

### 5.2 Verification

Participants described various strategies for verifying potentially inaccurate information. 55% of participants reported that they would try to corroborate the claim with multiple sources. As one participant reported, "I would Google the information and see how many other websites were posting the same information." 14% reported that they would verify the information by going directly to a trusted source, with some naming specific types of entities. For example, "I would go claim by claim in the article, cross-checking them with other sources that I do trust such as professional organizations, science outlets, and govt institutions." Only 13% stated that they would do further research into the source of the information or entity doing the fact-checking. A summary of participant responses is provided in Appendix H.

### 5.3 Trustworthiness of Entities

There was a significant difference for both topics in participants' ratings of how trustworthy they found the referenced entity with regards to presenting or assessing news. For the anti-Muslim headlines, the Collaboratory Against Hate received the highest ratings of trust and Facebook the least (see Figure 4(a)), while for COVID-19 headlines WHO received the highest ratings of trust and NAE the least (see Figure 4(b)).

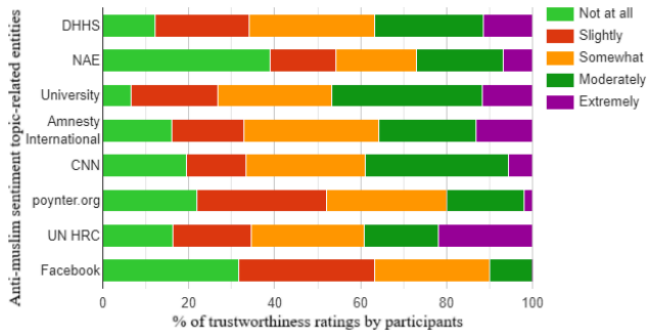

(a) Participants' level of trust in the entities referenced in headlines expressing anti-Muslim sentiment.

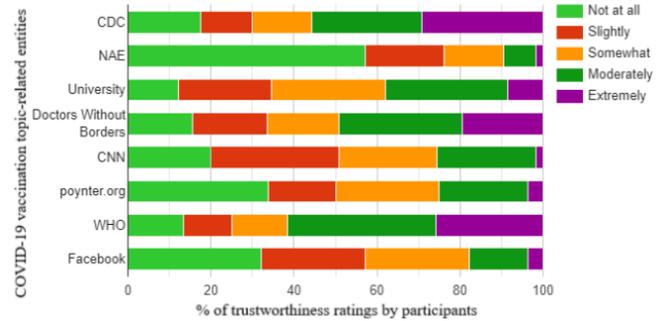

(b) Participants' level of trust in the entities referenced in COVID-19 headlines.

Figure 4: Participants' sentiment and pre-existing belief towards anti-Muslim sentiment and COVID-19 related statements in a 5-point likert scale ranging from -2 to 2

Similar to ratings of helpfulness, participants' rating in trustworthiness in certain entities significantly varied with age, political preference, and familiarity. Across age groups, CNN (p= 0.01) and the UN Human Rights Council (p= 0.009) had significant differences for anti-Muslim headlines and CDC (p= 0.05) and WHO (p= 0.003) had significant differences for COVID-19 headlines. Participants older than 45 rated the assessments by CNN and the UN Human Rights Council as less trustworthy than younger age groups, while those 18-30 had more trust in the CDC and WHO compared to older participants. Similar to helpfulness ratings, liberals had significantly less trust in the NAE, while conservatives reported significantly less trust in all other entities except Poynter.org and Facebook for both anti-Muslim and COVID-19 headlines. Familiarity also significantly impacted ratings of trust for five entities (UN Human Rights Council, DHHS, Amnesty International, NAE, and CNN) for anti-Muslim headlines and two entities (Doctors Without Borders and WHO) for COVID-19 headlines.

We also observed an overall difference in perceptions of bias across entities for both anti-Muslim headlines (p < 0.001) and COVID-19 headlines (p < 0.001). The Collaboratory Against Hate was perceived the least biased for the anti-Muslim headlines, with 42% of participants selecting that it "tends to deal fairly with all sides" while CNN was rated as the most biased with 64% selecting that it "tends to favor one side." For the COVID-19 headlines, both CDC and WHO were rated the least biased with 49% of participants indicating that these entities were fair with all sides while CNN was again rated the most biased with 67% of participants reporting that it favors one side. As with helpfulness and trustworthiness, we saw significant differences in perceptions of bias of some entities across age groups and political preferences.

Participants were asked to think of an entity they trust to provide accurate information and describe why they trust that entity. 35% reported that they believe the entity to be unbiased and factual in its reporting (e.g., "They are not biased towards either side and just share facts"). Nearly a third 27% considered the entity's history of providing accurate information or reputation (e.g. "I trust them due to its long history and extensive track record"). Some participants (15%) explained that the entity was able to provide references for information or that they trusted the references used in the reporting (e.g., "They provide a lot of sources in their articles."). CNN, NPR, and the New York Times were most frequently mentioned as trusted news entities. 17% of participants mentioned other reasons, such as trusting sources that they believe have trained professionals. Others also mentioned just relying on their own discretion of the entity's professionalism.

Participants were also asked the converse, to explain why they would not trust an entity to provide accurate

**COLLABORATORY AGAINST HATE** | **RESEARCH AND ACTION CENTER**

information. Similarly, 51% of participants did not trust entities they perceived as biased in their reporting or following a certain agenda (e.g., "I don't trust them because every piece of content they produce only focuses on one side/group and never their own"). Further, 24% stated that they did not trust entities that publish news they consider fake, while 9% did not trust that they did not provide references for their claims (e.g., "They spew lies that are not backed by anything"). Additionally, 9% described not trusting entities that used "clickbait" or published emotionally charged content (e.g., "They are emotionally evocative and motivated by prompting people to outrage"). 44% of participants named specific news entities they did not trust, with Fox News and CNN being the most frequently mentioned.

Appendix I provides a detailed summary of participants' responses related to trustworthiness of entities.

### 5.4 Sharing Intentions

We also asked participants how likely they would share the headlines they evaluated. Almost three-quarters of the ratings (73%) reported they would be extremely unlikely to share, while another 14% were somewhat unlikely. Only 6% reported they would be somewhat or extremely likely to share the content, while 7% said they would be neither likely nor unlikely.

Participants were asked to explain why they would or would not share the headline displayed. 32% of responses that indicated they would be extremely or somewhat unlikely to share explained that it was because they did not want to share inaccurate content. Almost a quarter (24%) of participants indicated they typically do not share news on social media, while 20% indicated they were not interested in sharing about these topics. 10% were not interested in sharing due to other people, where the participant considered the feedback of their social media friends, including whether or not it would interest them or cause conflict. Interestingly, 7% of participants indicated not trusting the fact-checking entity as a reason for not sharing, suggesting that some participants may have misinterpreted the content label as the entity *verifying* the claims in the content. Others expressed different reasons for not sharing, including lack of knowledge about the topic, and disagreeing with the headline.

## 6. Discussion

In our online study exploring perceptions of fact-checking entities across two misinformation topics, results showed variations based on age, political preference, and familiarity. Regardless of political stance, the assessment by Poynter.org was consistently seen as helpful and trustworthy. Although participants had differing views on the entities, most reported they would be unlikely to share the evaluated headlines. We discuss these findings and offer recommendations for addressing misinformation.

### 6.1 Limitations

Our use of self-reported data may have its limitations due to social desirability bias [20], which is the tendency for people to present themselves in a more positive manner according to social norms. Participants might have been inclined to report less sharing, given the stigma associated with sharing fake news. Further, some may have been more positive in their perceptions of certain entities (such as the university conducting the study). However, given the wide variation observed in participant perceptions, the impact of this bias appears to be minimal. Our results are also limited to a US population and perceptions of different categories of entities may vary across cultural contexts.

### 6.2 Effectiveness of Existing Misinformation Interventions

While not our main focus, our results hint at the effectiveness of current fact-checking measures on social media. The reported unlikelihood of participants sharing the content evaluated suggests such labels might deter sharing inaccurate information. However, the impact of fact-checking organizations might be limited if users are not familiar with them. Meta employs the International Fact-Checking Network from Poynter.org, also supported by Google and YouTube [22], which include many small, regional entities. Given our results related to familiarity with an organization, it is possible that assessments from lesser-known organizations may be less effective in preventing the sharing of false content than those from established well-reputed entities.

We based the design of the content participants' evaluated in the study on Facebook's content label warning, which is similar to those used on other platforms. Our results indicate that a small percentage of our participants may have had misinterpretations about the fact-check provided, specifically that the entity cited has fact-checked the content to be true. Though such misconceptions did not influence our overall results, it suggests that platforms should further evaluate the wording of content labels to avoid incorrect misinterpretations and improve user understanding.

### 6.3 Recommendations for Misinformation Interventions

Our findings underscore the growing partisan divide in US news dissemination, evident in differing opinions on entities like WHO and CNN. Our results supplement previous findings by Flintham et al highlighting skepticism related to journalistic style [13] as participants mistrusted entities that they perceived as biased or one-sided. While perceptions of most entities we studied varied based on political preference, dedicated fact-checkers like Poynter.org seem to be viewed as neutral or nonpartisan, given the lack of observed significant differences between conservatives and liberals. Further, our results suggest that, in some contexts, university centers may also be viewed as unbiased. Further research is needed to explore whether perceptions related to fact-checking organizations and university centers generalize beyond the specific entities explored in this study.

Considering that identifying multiple sources was the most common strategy our participants reported to verify potentially inaccurate claims, another mitigation against perceived bias in accuracy assessments would be to provide references to multiple entities reaching the same conclusion. These entities could include news entities or organizations with biases on both sides of the political spectrum [1], as long as they are reliable sources of information.

Building on prior recommendations by Bradshaw et al. [3], an additional strategy to consider in the design of misinformation interventions is the possibility of customized interventions. Given that social media algorithms already serve users content based on what they have learned about the user, they could potentially use the same data to provide fact-checking entities that are more likely to resonate with a particular user. Specifically, age and political preference were found to have significant impact on perceptions of entities; both are likely already demographic factors social media platforms use to target content.

---

## 7. Conclusion

We conducted a 655-participant online study to explore perceptions of different fact-checking entities across two topics of misinformation. We found that the perceived helpfulness of accuracy assessments and trust in the entities we evaluated varied greatly and largely aligned with participants' political preferences. However, the presence of the fact-checking label appeared to impact participants' sharing intentions. Our results provide insight into the effectiveness of existing misinformation interventions and recommendations for the design of such future interventions.

**COLLABORATORY AGAINST HATE** | **RESEARCH AND ACTION CENTER**

## Appendix A: Survey Questions

Imagine that you saw the following posts on a social media platform. Please answer the following questions with regards to the displayed posts. Note that you should not reveal any private or personally-identifiable information about yourself OR others in answers to your open-ended questions. [Let's consider "CNN" as a randomly chosen fact checking entity.]

### Investigating entities

How helpful is the accuracy assessment provided by CNN

**COLLABORATORY AGAINST HATE** | **RESEARCH AND ACTION CENTER**

for this headline?

- Not at all
- Slightly
- Somewhat
- Moderately
- Extremely

Why do you believe the assessment provided by CNN for this news article was helpful/not helpful? (follow up question)

If you were to see the above article on social media, how likely would you be to share it?

- Extremely unlikely
- Somewhat unlikely
- Neither likely nor unlikely
- Somewhat likely
- Extremely likely

Why would you share/not share the article on social media? (follow up question)

How trustworthy would you rate CNN with regards to assessing or presenting news?

- Not at all
- Slightly
- Somewhat
- Moderately
- Extremely

In assessing or presenting news, do you think that CNN deals fairly with all sides, or does it tend to favor one side?

- CNN tends to deal fairly with all sides.
- CNN tends to favor one side.
- Not sure

Which of the following best describes how familiar you are with the work of CNN?

- I've never heard of CNN before.
- I have heard of CNN but do not know what it does.
- I have read a few news articles on or about CNN and am somewhat familiar with its work.
- I read news articles on or about CNN frequently and am very familiar with its work.

Have you seen or heard about the above headline before?

- Yes
- No
- Unsure

## Trust Practices and Judgements of News Accuracy

Suppose you saw an article similar to the ones in this survey and you were unsure about its accuracy. If you were curious about the accuracy, how would you go about verifying the claim?

Please rate your confidence in the following categories of entities separately.

1. National news organizations
2. Local news organizations
3. Fact-checking organizations
4. Social networking sites (e.g., Facebook, Twitter)
5. The government
6. NGOs or Non-profit organizations
7. Universities

Rating:

- None at all
- Very much
- A moderate amount
- Quite a lot
- A great deal

Think of an entity (e.g., news publisher, nonprofit organization, public figure, government agency) that you trust to provide accurate information. Why do you trust them?

Think of an entity (e.g., news publisher, nonprofit organization, public figure, government agency) that you do not trust to provide accurate information. Why do you not trust them?

## Demographic and Background Information

Generally speaking, how often do you pay attention to information about politics and public affairs?

- None at all
- Sometimes
- About half of the time
- Most of the time
- All of the time

Which of the following best describes your political preference?

- Strongly Liberal
- Somewhat Liberal
- Moderate
- Somewhat Conservative
- Strongly Conservative
- Prefer not to say

Select the statement that best describes your belief towards the following [answered as "I strongly believe that", "I kind of believe that", "I'm not sure I believe that", "I don't really believe that", or "I strongly don't believe that"]:

**CS1** Bill Gates, George Soros, or some other powerful person is behind coronavirus

**CS2** The coronavirus vaccine has microchips in it to secretly track people who get vaccinated.

**CS3** The coronavirus vaccine may cause infertility so people who might want to have children in the future shouldn't get it.

**CS4** Wearing a cloth face covering/mask in public is effective in slowing the spread of the coronavirus.

**CS5** Staying at least 6 feet from people not in your household is effective in slowing the spread of the coronavirus.

Select the statement that best describes your belief towards the following [answered as "I strongly believe that", "I kind of believe that", "I'm not sure I believe that", "I don't really believe that", or "I strongly don't believe that"]:

**AS1** Most Muslims living in my country are more prone to violence than other people.

**AS2** Most Muslims living in my country discriminate against women.

**AS3** Most Muslims living in my country are hostile to my country.

**AS4** Most Muslims living in my country are less civilized than other people.

**AS5** Most Muslims living in my country are partially responsible for acts of violence carried out by other Muslims.

Do you belong to an ethnic or religious minority group in your country of residence?

- No
- Yes
- Prefer not to say

What is your age? (Enter 0 if you prefer not to say.)

What is your gender?

- Male
- Female
- Non-binary/third-gender
- Prefer to self identify
- Prefer not to say

What is the highest level of school you have completed or the highest degree you have received?

- Less than high school degree
- High school graduate (high school diploma or equivalent
- including GED)
- Some college but no degree
- Associate degree in college (2-year)
- Bachelor's degree in college (4-year)
- Master's degree
- Doctoral degree
- Professional degree (JD, MD)
- Prefer not to say

Please indicate the answer that includes your entire household income in (previous year) before taxes.

- Less than $10,000
- $10,000 - $19,999
- $20,000 - $29,999
- $30,000 - $39,999
- $40,000 - $49,999
- $50,000 - $59,999
- $60,000 - $69,999
- $70,000 - $79,999
- $80,000 - $89,999
- $90,000 - $99,999
- $100,000 - $149,999
- More than $150,000
- Prefer not to say

**COLLABORATORY AGAINST HATE** | **RESEARCH AND ACTION CENTER**

## Appendix B: The 2017 US Results of World Values Survey

| Entities | Not at all | Not very much | Quite a lot | A great deal |
|---|---|---|---|---|
| | % of Participants | | | |
| The UN | 15.4 | 39.2 | 35.8 | 7.9 |
| WHO | 9.7 | 36.3 | 39.5 | 12.1 |
| The Press | 23.8 | 45.9 | 24.2 | 5.3 |
| Television | 16.9 | 59.6 | 19.5 | 2.9 |
| The government | 29 | 36.7 | 25.1 | 8.3 |
| NGOs | 5 | 32 | 50 | 11.8 |
| Universities | 9 | 35.8 | 45.2 | 8.7 |
| Major Company | 13.8 | 53.7 | 28.9 | 2.3 |
| Religion | 11.2 | 34.5 | 40.1 | 13 |

Table 4: The 2017 US results of World Values Survey

## Appendix C: Helpfulness of Entities Codebook

| Code | Definition | Example |
|---|---|---|
| entity - positive | Participant says something about the entity in a positive light | Maybe used as propaganda but since it's the UN it may have some inaccuracy in order to persuade the viewers towards a specific perspective. |
| entity - negative | Participant says something about the entity in a negative light | It's biased. Another religion fact checking another religion's action just seems biased to me. Needs to be someone who is neutral. |
| not enough info | Participant says there are details missing that prevent the assessment from being helpful | I'm not sure who actually fact checked what and what was fact checked |
| fact-check important | Participant says it's helpful to have inaccuracies flagged, but doesn't specifically refer to an entity | It's important for factual inaccuracies to be pointed out to people |
| fact-check not important | Participant says it's not helpful to have inaccuracies flagged or that they are biased, but doesn't specifically refer to an entity | Who are independent fact checkers, maybe biased? |

| not familiar with entity | participant mentions that they are not familiar with the entity sharing the news | I'm not familiar with Amnesty International. |
|---|---|---|
| not familiar with topic | participant mentions not interested or aware of topic of the news information being shared | I do not know much about soccer and wouldn't know whether this was true or not so having a credible authority helps me make that decision. |
| very familiar with topic | participant says they know enough about the news and are able to identify inaccuracies themselves, or they have a subjective opinion related to the topic | The headline is ridiculous enough that I would trust any news source that found inaccuracies. |

Table 5: Helpfulness of Entities Codebook

## Appendix D: Trustworthiness of Entities Codebook

### Trust Question

| Code | Definition | Example |
|---|---|---|
| Entity's Impartiality (No Agenda/Bias) | Participant says that they believe the entity is unbiased/unopinionated | Because the way in which they present information is as straightforward and un-opinionated as I've found. |
| Entity doesn't spread misinformation | Participant mentions that the entity is known to lie or spread misinformation, or publishing opinion as fact. Focus is on the accuracy of the news reported | They seem to give good, accurate information most of the time and usually walk back statements if they're wrong. |
| Entity's references | Participants say they trust entities using first hand reports, knowledge, or skill. Focus is on the sources used to report news. | They provide local information directly from first hand sources or police reports |
| Entity's funding | Participant says that they trust publicly-funded entities over corporate interests | NPR, it is a publicly funded new source and does not have to skew their information to serve corporate interests. |
| Entity's reputation | Participants mention that they trust the entity based on it's reputation or previous proven record. | They've proven themselves with a long track record of facts. |
|  | Participant mentions knowing someone or experiencing something through the entity that built their trust | The US military because I was once a member and know their intelligence validation procedures. |
| Entity's support of personal beliefs | Participant trust entity that aligns with participant's personal views. | They have similar values to me |
| Trust | Participants mention that they trust the entity. Can be used for general commentary that doesn't mention a specific reason for trusting the entity. | Unless it's coming from my people they know who they are. Or individual research. I never believe that shit. I take it all with a grain of salt. |
| Don't Trust | Participant mentions they don't trust or fully trust any entity | I don't trust any of them |

**COLLABORATORY
AGAINST HATE** | **RESEARCH AND
ACTION CENTER**

| Other | Participant mentions other reasons | BBC news- its outside of the USA |
|---|---|---|

Table 6: Trust Question

### Don't Trust Question

| Code | Definition | Example |
|---|---|---|
| Entity's Agenda/Bias | Participant mentions that they do not trust entities that clearly have an agenda or are biased in their news presentation | infowar like sites, do not trust because of bias. |
| Entity's lack of source/reference | Participant mentions that they do not trust articles that do not present references, or present confusing or contradictory statements | Inability to provide source information. |
| Entity spreads misinformation | Participant mentions entity is known to lie or spread misinformation, or publishing opinion as fact | Because they've said things that turned out to be untrue |
| Entity's funding | Participant mentions they do not trust the entity's financial interests/funding | Because they follow the money. |
| Entity's emotional manipulation | Participant mentions the entity's strategy of presenting the news is mainly to manipulate people through inciting emotions | it incites emotions rather than critical thinking<br><br>They proved the only thing they care about is online clicks and trendiness. |
| Entity's reputation | Participant mentions that they do not trust the entity based on it's reputation or previous proven record | Fox news. For obvious reasons |
| Don't trust | Participant says they don't trust anyone | I don't trust any specific source. |
| Other | Participant mentions that they will trust entity | I will always trust and fully focus on them |

Table 7: Don't Trust Questions

### Appendix E: Sharing Intentions Codebook

| Code | Definition | Example |
|---|---|---|
| No - Due to other people | Participant considers the feedback of their social media friends, including whether or not it would interest them or cause conflict | I don't like to share things that make other people upset. |
| No - Lack of interest in news topic | Participant mentions that they are not interested in the topic of the news (separate from not interested in SHARING about the topic) | I'm not interested in this or sharing it. |
| No - Lack of knowledge about topic | Participant mentions that they do not have enough information about the topic or entity | I do not know enough about this topic to share my thoughts |

| No - Inaccuracy | Participant mentions that the reason they wouldn't share the news is because of the inaccuracy or need for further fact-checking | I would need to fact check it myself |
|---|---|---|
| No - Bias | Participant mentions that the reason they wouldn't share the news is because it's biased or intended to create outrage | It's probably highly biased. |
| No - Typically do not share news | Participant not interested in sharing news regardless of accuracy, or news related to these topics | I typically do not share posts. It's not really my thing.<br><br>I do not usually share articles as politically charged as this one is |
| No - Don't trust the entity | Participant mentions they will not share the article because they do not trust the entity (actually a misconception about the label) | I have a lot of Muslim friends and do not trust CNN and would not share anything reviewed by them.<br><br>WHO is a globalist agency and I won't do anything to bring them attention |
| No - They trust the entity | Participant mentions they will not share the article because they trust an entity that said the news is incorrect | If the WHO says it's incorrect, I won't be sharing it. |
| No - Disagree with news | Participant doesn't want to promote the news, or has an opposing opinion. Also if they express some subjective opinion either in-line or opposing the headline or fact-checking assessment. | I am not vaxed against Covid 19 and I would not share anything about it that would be hypocritical. |
| Yes - | Participant mentions that they will share the news | Very special and unique and willing to share |
| Other - | Participant mentions a personal perspective about sharing | I would take it with a grain of salt but it's just not that important to share. I think people can form their own options without me offering any assistance. |

Table 8: Sharing intentions codebook

## Appendix F: Verification Questions Codebook

| Code | Definition | Example |
|---|---|---|
| More Research on Topic | Participants mention that they would first do additional search (ex. Googling) to verify the information presented. | Googling and learning more about the thing in question myself. |
| More Research on Entity | Participants mention that they would first do additional search (ex. Googling) to compare entities history or current publication's coverage. | I would start with the review of the agency that flagged it, then branch out into independent research. |
| Verify Info through another Fact-checker | Participants mention that they would first check another fact-checking entity that they trust to see if it | I would google it to see if it was covered elsewhere on major news sites and then go to Snopes. |

COLLABORATORY AGAINST HATE | RESEARCH AND ACTION CENTER

| they trust | matches. | |
|---|---|---|
| Verify info through a specific entity | Participants mention that they would first check a specific fact-checking entity that they trust. | First I would Google it. Then I would look at the articles in regard to it and find one that is from a valid source like CNN or MSNBC. |
| Ignoring | Participants mention they wouldn't share but just ignore it or are not interested. | I wouldn't just ignore it. |
| Not sure | Participants say they wouldn't know if they will share it or not. | Not sure what I will do in this case |
| Other | Participants mention a specific strategy of identifying false information and how they prefer to verify it. | Search keywords within articles that are not opinionated and compare other sources. |
| Find other sources | Participants mention that they would check references or additional similar sources covering the news. | |

Table 9: Verification Questions Codebook

## Appendix G: Helpfulness Question Results

| | Headline 1 | | | | | |
|---|---|---|---|---|---|---|
| | Extremely | Moderately | Somewhat | Slightly | Not at all | |
| entity - positive | 14 | 13 | 6 | 3 | 0 | 36 |
| entity - negative | 0 | 0 | 11 | 12 | 23 | 46 |
| not enough info | 1 | 18 | 29 | 24 | 12 | 84 |
| fact-check important | 28 | 36 | 22 | 12 | 1 | 99 |
| fact-check not important | 0 | 0 | 2 | 2 | 9 | 13 |
| not familiar with entity | 0 | 2 | 4 | 1 | 8 | 15 |
| not familiar with topic | 0 | 0 | 1 | 2 | 3 | 6 |
| very familiar with topic | 4 | 4 | 8 | 3 | 2 | 21 |
| other | 1 | 5 | 5 | 4 | 0 | 15 |
| | 48 | 78 | 88 | 63 | 58 | 335 |

| | Headline 2 | | | | | |
|---|---|---|---|---|---|---|
| | Extremely | Moderately | Somewhat | Slightly | Not at all | |
| entity - positive | 13 | 15 | 10 | 5 | 0 | 43 |
| entity - negative | 0 | 3 | 5 | 5 | 25 | 38 |
| not enough info | 0 | 11 | 20 | 22 | 20 | 73 |

| | | | | | |
|---|---|---|---|---|---|
| fact-check important | 32 | 27 | 24 | 4 | 3 | 90 |
| fact-check not important | 0 | 0 | 2 | 4 | 9 | 15 |
| not familiar with entity | 0 | 3 | 2 | 3 | 4 | 12 |
| not familiar with topic | 0 | 1 | 4 | 3 | 6 | 14 |
| very familiar with topic | 3 | 3 | 8 | 5 | 10 | 29 |
| other | 1 | 4 | 5 | 1 | 3 | 14 |
| | 49 | 67 | 80 | 52 | 80 | 328 |

| | Headline 3 | | | | | |
|---|---|---|---|---|---|---|
| | Extremely | Moderately | Somewhat | Slightly | Not at all | |
| entity - positive | 24 | 32 | 12 | 1 | 0 | 69 |
| entity - negative | 0 | 1 | 10 | 13 | 38 | 62 |
| not enough info | 0 | 15 | 21 | 21 | 17 | 74 |
| fact-check important | 20 | 18 | 15 | 3 | 4 | 60 |
| fact-check not important | 0 | 2 | 2 | 7 | 11 | 22 |
| not familiar with entity | 0 | 2 | 1 | 8 | 4 | 15 |
| not familiar with topic | 0 | 0 | 0 | 0 | 0 | 0 |
| very familiar with topic | 3 | 2 | 3 | 4 | 8 | 20 |
| other | 2 | 2 | 6 | 6 | 3 | 19 |
| | 49 | 74 | 70 | 63 | 85 | 341 |

| | Headline 4 | | | | | |
|---|---|---|---|---|---|---|
| | Extremely | Moderately | Somewhat | Slightly | Not at all | |
| entity - positive | 30 | 18 | 12 | 3 | 0 | 63 |
| entity - negative | 0 | 2 | 5 | 11 | 31 | 49 |
| not enough info | 0 | 15 | 18 | 13 | 11 | 57 |
| fact-check important | 36 | 27 | 22 | 13 | 0 | 98 |
| fact-check not important | 0 | 3 | 1 | 0 | 8 | 12 |
| not familiar with entity | 0 | 2 | 2 | 2 | 6 | 12 |
| not familiar with topic | 0 | 0 | 0 | 0 | 0 | 0 |
| very familiar with topic | 1 | 3 | 8 | 5 | 5 | 22 |
| other | 2 | 9 | 3 | 3 | 3 | 20 |
| | 69 | 79 | 71 | 50 | 64 | 333 |

**COLLABORATORY AGAINST HATE** | **RESEARCH AND ACTION CENTER**

| | | Extremely | Moderately | Somewhat | Slightly | Not at all | Total |
|---|---|---|---|---|---|---|---|
| Total Responses | Headline 1 | 48 | 78 | 88 | 63 | 58 | 335 |
| | Headline 2 | 49 | 67 | 80 | 52 | 80 | 328 |
| | Headline 3 | 49 | 74 | 70 | 63 | 85 | 341 |
| | Headline 4 | 69 | 79 | 71 | 50 | 64 | 333 |
| | **Total** | **215** | **298** | **309** | **228** | **287** | 1337 |

## Appendix H: Verification Question Results

| Code | Responses |
|---|---|
| More Research on Topic | 86 |
| More Research on Entity | 36 |
| Verify Info through another Fact-checker they trust | 92 |
| Verify info through a specific entity | 59 |
| Ignoring | 8 |
| Not sure | 10 |
| Other | 69 |
| Find other sources | 359 |
| | |
| **Total** | **654** |

## Appendix I: Trustworthiness of Entities Question Results

### Trust

| Code | Responses |
|---|---|
| Entity's Impartiality (No Agenda/Bias) | **227** |
| Entity doesn't spread misinformation | **60** |
| Entity's references | **97** |
| Entity's funding | **23** |
| Entity's reputation | **175** |
| Entity's support of personal beliefs | **21** |
| Trust | **4** |
| Don't Trust | **29** |
| Other | **108** |
| | |
| **Total responses** | **654** |

## Don't Trust

| Code | Responses |
|---|---:|
| Entity's Agenda/Bias | **335** |
| Entity's lack of source/reference | **57** |
| Entity spreads misinformation | **156** |
| Entity's funding | **43** |
| Entity's emotional manipulation | **56** |
| Entity's reputation | **38** |
| Don't trust | **11** |
| Intentions Other | **41** |
| | |
| **Total responses** | **654** |

## Appendix J: Sharing Intentions Question Results

| Headline 1 | | | | | | |
|---|---|---|---|---|---|---|
| | Extremely Unlikely | Somewhat unlikely | Neither likely or unlikely | Somewhat likely | Extremely likely | Total |
| No - Due to other people | 17 | 52 | 2 | 1 | 0 | **72** |
| No - Lack of interest in news topic | 57 | 12 | 3 | 0 | 0 | **72** |
| No - Lack of knowledge about topic | 15 | 1 | 1 | 0 | 0 | **17** |
| No - Inaccuracy | 81 | 16 | 2 | 1 | 0 | **100** |
| No - Bias | 15 | 1 | 0 | 0 | 0 | **16** |
| No - Typically do not share news | 67 | 14 | 6 | 0 | 0 | **87** |
| No - Don't trust the entity | 19 | 3 | 0 | 0 | 0 | **22** |

| | Extremely Unlikely | Somewhat unlikely | Neither likely or unlikely | Somewhat likely | Extremely likely | Total |
|---|---|---|---|---|---|---|
| No - They trust the entity | 5 | 1 | 0 | 1 | 0 | **7** |
| No - Disagree with news | 7 | 0 | 0 | 0 | 0 | **7** |
| Yes - | 0 | 0 | 1 | 8 | 4 | **9** |
| Other - | 7 | 4 | 3 | 1 | 1 | **15** |
| | | | | | | |
| **Total responses** | **247** | **52** | **16** | **10** | **5** | |

| Headline 2 | | | | | | |
|---|---|---|---|---|---|---|
| | Extremely Unlikely | Somewhat unlikely | Neither likely or unlikely | Somewhat likely | Extremely likely | Total |
| No - Due to other people | 14 | 4 | 1 | 0 | 0 | **19** |
| No - Lack of interest in news topic | 86 | 16 | 1 | 0 | 0 | **103** |
| No - Lack of knowledge about topic | 10 | 5 | 0 | 0 | 0 | **15** |
| No - Inaccuracy | 87 | 10 | 4 | 2 | 0 | **103** |
| No - Bias | 13 | 2 | 0 | 0 | 0 | **15** |
| No - Typically do not share news | 59 | 10 | 3 | 0 | 0 | **72** |
| No - Don't trust the entity | 13 | 0 | 0 | 0 | 0 | **13** |
| No - They trust the entity | 4 | 0 | 0 | 1 | 0 | **5** |
| No - Disagree with news | 10 | 1 | 0 | 1 | 0 | **12** |
| Yes | 0 | 0 | 0 | 2 | 1 | **2** |
| Other | 14 | 3 | 2 | 0 | 3 | 19 |
| | | | | | | |
| **Total responses** | **256** | **47** | **11** | **7** | **3** | |

| Headline 3 | | | | | | |
|---|---|---|---|---|---|---|
| | Extremely Unlikely | Somewhat unlikely | Neither likely or unlikely | Somewhat likely | Extremely likely | Total |
| No - Due to other people | 6 | 5 | 2 | 1 | 0 | **14** |
| No - Lack of interest in news topic | 23 | 3 | 4 | 0 | 0 | **30** |
| No - Lack of knowledge about topic | 0 | 0 | 0 | 0 | 0 | **0** |
| No - Inaccuracy | 58 | 9 | 5 | 0 | 0 | **72** |
| No - Bias | 4 | 0 | 0 | 0 | 0 | **4** |
| No - Typically do not share news | 52 | 9 | 8 | 0 | 0 | **69** |
| No - Don't trust the entity | 25 | 1 | 0 | 0 | 0 | **26** |
| No - They trust the entity | 6 | 0 | 0 | 0 | 0 | **6** |
| No - Disagree with news | 9 | 2 | 0 | 1 | 0 | **12** |
| Yes | 0 | 0 | 0 | 12 | 6 | **12** |
| Other | 18 | 3 | 11 | 2 | 1 | **34** |
| | | | | | | |
| **Total responses** | **236** | **40** | **31** | **15** | **7** | |

| Headline 4 | | | | | | |
|---|---|---|---|---|---|---|
| | Extremely Unlikely | Somewhat unlikely | Neither likely or unlikely | Somewhat likely | Extremely likely | **Total** |
| No - Due to other people | 13 | 6 | 6 | 1 | 0 | **26** |
| No - Lack of interest in news topic | 24 | 6 | 3 | 0 | 0 | **33** |
| No - Lack of knowledge about topic | 5 | 0 | 1 | 0 | 0 | **6** |
| No - Inaccuracy | 82 | 19 | 4 | 2 | 1 | **107** |
| No - Bias | 5 | 0 | 0 | 0 | 0 | **5** |
| No - Typically do not share news | 50 | 9 | 11 | 2 | 0 | **72** |
| No - Don't trust the entity | 19 | 2 | 3 | 0 | 0 | **24** |
| No - They trust the entity | 7 | 1 | 0 | 2 | 0 | **10** |

**COLLABORATORY AGAINST HATE** | **RESEARCH AND ACTION CENTER**

| | | | | | | |
|---|---|---|---|---|---|---|
| No - Disagree with news | 15 | 3 | 3 | 4 | 0 | **25** |
| Yes | 0 | 0 | 2 | 16 | 2 | **18** |
| Other | 13 | 2 | 8 | 2 | 0 | **25** |
| | | | | | | |
| **Total responses** | **211** | **47** | **37** | **27** | **3** | |

| | Extremely Unlikely | Somewhat unlikely | Neither likely or unlikely | Somewhat likely | Extremely likely | |
|---|---|---|---|---|---|---|
| Total responses | 247 | 52 | 16 | 10 | 5 | |
| | 256 | 47 | 11 | 7 | 3 | |
| | 236 | 40 | 31 | 15 | 7 | |
| | 211 | 47 | 37 | 27 | 3 | |
| | **950** | **186** | **95** | **59** | **18** | **Total:1308** |

Funding for this project was provided by the Collaboratory Against Hate. We would like to thank Dr. Geoff Kaufman, Likhitha Chintareddy, Hamsini Ravishankar, and other colleagues at CMU who provided input regarding the study design. Additionally, we would like to acknowledge the Fatima Fellowship Program for the opportunity to conduct this research.